\documentclass[final,5p,times,twocolumn,sort&compress]{elsarticle}
\usepackage{amsmath,amssymb,amsfonts,graphicx,slashed,citesort,dsfont}
\usepackage[usenames]{color}
\usepackage{ulem}
\usepackage{nicefrac} 
\newcommand{\non}{\nonumber \\}
\newcommand{\be}{\begin{equation}}
\newcommand{\ee}{\end{equation}}
\newcommand{\ba}{\begin{eqnarray}}
\newcommand{\ea}{\end{eqnarray}}
\let\oldhat\hat
\renewcommand{\vec}[1]{\boldsymbol{#1}}
\renewcommand{\hat}[1]{\oldhat{\boldsymbol{#1}}}
\allowdisplaybreaks[1]
\begin{document}
%%%%%%%%%%%%%%%%%%%%%%%%%%%%%%%%%%%%%%%%%%%%%%%%%%%%%%%%%%%%%%%%%%%%%%%%%%%%%

\begin{frontmatter}
\title{Finite volume effects and quark mass dependence of the $N(1535)$ and $N(1650)$}

\author[Bonn]{Michael D\"oring}
\author[Bonn]{Maxim~Mai}
\author[Bonn,Julich]{Ulf-G.~Mei{\ss}ner}

\address[Bonn]{Helmholtz--Institut f\"ur Strahlen- und Kernphysik (Theorie) 
   and Bethe Center for Theoretical Physics, Universit\"at Bonn, D-53115 Bonn, Germany}
\address[Julich]{Institut f\"ur Kernphysik (IKP-3), Institute for Advanced
  Simulation (IAS-4), J\"ulich Center for Hadron Physics and JARA-HPC, 
  Forschungszentrum J\"ulich, D-52425  J\"ulich, Germany}

\begin{abstract} 
For resonances decaying in a finite volume, the simple identification of state
and eigenvalue is lost. The extraction of the scattering
amplitude is a major challenge as we demonstrate by extrapolating the physical
$S_{11}$ amplitude of pion-nucleon scattering
to the finite volume and unphysical quark
masses, using a unitarized chiral framework including all next-to-leading
order contact terms. We show that the pole movement of the resonances
$N(1535)1/2^-$ and $N(1650)1/2^-$ with varying quark masses is non-trivial. In
addition, there are several strongly coupled $S$-wave thresholds
that induce a similar avoided level crossing as narrow resonances. The level spectrum is predicted for two typical lattice setups, and ways to
extract the amplitude from upcoming lattice data are discussed.  
\end{abstract}


\end{frontmatter}

%%%%%%%%%%%%%%%%%%%%%%%%%%%%%%%%%%%%%%%%%%%%%%%%%%%%%%%%%%%%%%%%%%%%%%%%%%%%%

\section{Introduction}
Pion-nucleon scattering has traditionally been the premier reaction to study the resonance excitations of the nucleon. In particular, in the
$S_{11}$ partial wave one finds two close-by resonances at 1535 and 1650~MeV, which overlap within their widths of about 100~MeV. It was pointed
out early in the framework of unitarized coupled-channel chiral perturbation theory \cite{Kaiser:1995cy} that the $N(1535)1/2^-$ might not be a
three-quark resonance, but is rather generated by strong channel couplings with a dominant $K\Sigma - K\Lambda$ component in its wave function.
This analysis was extended in Ref.~\cite{Inoue:2001ip}, where within certain approximations the effects of 3-body $\pi\pi N$ channels were also
included. Further progress was made in  Ref.~\cite{Nieves:2001wt}, where the $S_{11}$ phase shift was fitted from threshold to about $W=\sqrt{s}
\simeq 2\,$GeV together with cross section data for $\pi^- p\to \eta n$ and $\pi^- p\to K^0 \Lambda$ in the respective threshold regions. More
recently, it was pointed out in a state-of-the-art unitary meson-exchange model~\cite{Doring:2009yv} that there is indeed  strong resonance
interference between the two $S_{11}$ resonances, as each of these resonances provides an energy-dependent background in the region of the
other. In Ref.~\cite{Bruns:2010sv} the coupled-channel problem in the
$J^P=1/2^-$ (with $J$ the spin and $P$ the parity)
sector was addressed, for the first time, using the full
off-shell Bethe-Salpeter equation and all contact terms of the leading and
next-to-leading order (NLO) in the chiral expansion of the meson-baryon interaction. Remarkably, not
only the $N(1535)1/2^-$ emerged from the meson-baryon dynamics, but also the $N(1650)1/2^-$ could be predicted without being included in the
fit.

Another source of experimental information on the $J^P=1/2^-$ and other resonances is provided by the dedicated baryon resonance programs at
ELSA, MAMI and Jefferson Lab~\cite{Denizli:2007tq,McNicoll:2010qk}. On the theoretical side, the concept of dynamical resonance generation has been
investigated
in~\cite{Mai:2012wy,Ruic:2011wf,Jido:2007sm,Doring:2009uc,Doring:2009qr} 
comparing with the extracted multipoles and helicity amplitudes from the
SAID and MAID analyses~\cite{Workman:2011hi,Workman:2012jf,Chen:2012yv,Drechsel:2007if}. Of particular interest is the gauge invariant scheme
developed for the full off-shell Bethe-Salpeter equation~\cite{Borasoy:2007ku} that has been applied to pion and eta
photoproduction~\cite{Mai:2012wy,Ruic:2011wf}.

Finally, lattice gauge simulations have rapidly evolved and the spectrum of
excited baryons starts to become accessible, in particular also for the
$J^P=1/2^-$ sector~\cite{Lang:2012db,Schiel:2011av,Bulava:2010yg,Basak:2007kj,Engel:2010my,Mathur:2003zf}. As quark masses come closer to the
physical limit, finite volume effects dominate the lattice spectrum. Further, as resonances start to decay their signal on the lattice is lost. Still,
L\"uscher has shown how to model-independently extract phase shifts from
lattice levels~\cite{luscher,Luscher:1990ux} (see also \cite{Wiese:1988qy}).
For example, Lang and
Verduci recently provided such levels above threshold, for the first time in the $J^P=1/2^-$ sector~\cite{Lang:2012db}. L\"uscher's method can
be combined with effective field theory to study baryonic resonances and their width in the finite volume~\cite{Bernard:2007cm,Bernard:2008ax}.
The extension to coupled channels has been pioneered in
Ref.~\cite{Lage:2009zv} (see also~\cite{Liu:2005kr})
and further applied to excited
mesons~\cite{Bernard:2010fp,Doring:2011vk,Li:2012bi}. For a different
approach, see Ref.~\cite{Hall:2012wz}.

Using these techniques in combination with the unitarized chiral approach of Ref.~\cite{Bruns:2010sv}, we predict in this Letter the
finite-volume level spectrum of the $S_{11}$ partial wave, extrapolated to
unphysical quark masses. 
In addition, we test the hypothesis that the
hidden-strangeness $KY$ channels provide the crucial dynamics for the resonance generation by applying twisted boundary conditions for the
strange quark. As we will show, the interplay between thresholds and
resonances is very intricate and need to be accounted for in any
extraction of resonance properties in this partial wave (and for other processes
that exhibit similar properties).

%%%%%%%%%%%%%%%%%%%%%%%%%%%%%%%%%%%%%%%%%%%%%%%%%%%%%%%%%%%%%%%%%%%%%%%%%%%%%

\section{Framework}

%\subsection{\textbf{$\boldsymbol{\pi N}$ scattering in the infinite volume}}
\subsection{\textbf{Pion-nucleon scattering in the infinite volume}}

In the present work we rely on the model for the description of meson-baryon scattering in the first and second resonance region as developed in
Ref.~\cite{Bruns:2010sv}. There, the Bethe-Salpeter equation has been solved  including the full off-shell dependence of the chiral potential.
The latter has been chosen to consist of all local  terms of first and second chiral order, omitting, however, the one--baryon exchange graphs
from the beginning. Including two-body channels with quantum numbers of the pion-nucleon system, the model describes the $S_{11}$ partial wave
rather well up to quite high energies, i.e. $W\lesssim1800$~MeV. In particular, this framework allows for a dynamical generation of both
negative-parity nucleonic resonances, the $N(1535)1/2^-$ and $N(1650)1/2^-$.

For two-particle scattering, we denote the in- and out-going meson momenta by $q_1$ and $q_2$, respectively. The overall four-momentum is
$p=q_1+p_1=q_2+p_2$, where $p_1$ and $p_2$ are the momenta of in- and outgoing baryon, respectively. For the unitary meson-baryon scattering
amplitude $T(q_2, q_1; p)$ and the potential $V(q_2, q_1; p)$, the Bethe-Salpeter integral equation reads in $d$ dimensions
\begin{align}
 T({q}_2, {q}_1; p)&=   V(q_2, q_1; p)\label{BSE2}\\
		   &   + i\int\frac{d^d \ell}{(2\pi)^d} \frac{V({q}_2, {\ell}; p)\,(\slashed{p}-\slashed{\ell}+m)\, 
		   T({\ell}, {q}_1; p)}{({\ell^2-M^2+i\epsilon})((p-\ell)^2-m^2+i\epsilon)}~,\nonumber
\end{align}
where $m$ and $M$ denote the mass of baryon and meson, respectively. This equation has to be understood as a matrix equation in channel space, and the channel
space is constructed from a certain number of the allowed combinations of one ground-state octet meson and one ground-state octet baryon. For isospin $I=1/2$
and strangeness $S=0$ the channels are $\pi N$, $\eta N$, $K\Lambda$, and $K\Sigma$. The propagator is diagonal in channel space.

By maintaining the full off-shell dependence, the identification of every term of Eq.~(\ref{BSE2}) with Feynman diagrams is ensured, which for
instance allows for the construction of a gauge invariant photoproduction amplitude~\cite{Mai:2012wy} in a very natural way. However, in a
finite volume the Passarino-Veltmann reduction utilized for the solution of Eq.~(\ref{BSE2}) in Refs.~\cite{Bruns:2010sv,Mai:2012wy} is a-priori
no longer applicable. To overcome this complication we set all tadpole integrals to zero in this solution, which puts for instance the potential
$V$ on the two-particle mass shell. This simplifies Eq.~(\ref{BSE2}) to the following algebraic equation
\begin{align}
 T^{\rm on}= V^{\rm on} + V^{\rm on}G T^{\rm on}~,%\nonumber
\end{align}
where all elements are again matrices in channel space and the remaining loop function $G$ reads
\begin{align}
G&:= i\mathop{\int} \frac{d^d\ell }{(2\pi)^d} \frac{\slashed{p}-\slashed{\ell}+m}{(\ell^2- M^2+i\epsilon)((p-\ell)^2- m^2+i\epsilon)}\non
	   &~= \Big(\slashed{p}\frac{p^2-M^2+m^2}{2 p^2}+m\Big)\,I_{MB}~.
\label{scalar}	   
\end{align}
Here $I_{MB}$ denotes the scalar one-meson-one-baryon loop integral. The factor in parenthesis in the second line yields 
\begin{align}
(\ldots)=2m+\vec\gamma\cdot\vec p_{\rm cms}~,
\end{align}
where $\vec p_{\rm cms}$ denotes the baryon three-momentum in the c.m. frame. This demonstrates the difference to another on-shell scheme that
is widely used in the literature (see, e.g., Refs.~\cite{Inoue:2001ip,Jido:2007sm}) in which the term $\vec\gamma\cdot\vec p_{{\rm cms}}$ is
omitted.

The renormalization of loop divergences in non-perturbative frameworks is
known to be complicated. Relying on the arguments given in
Refs.~\cite{Bruns:2010sv,Mai:2012wy}, we utilize dimensional regularization,
applying the usual $\overline{\rm MS}$ scheme. The
finite part of scalar loop integral reads in four space-time dimensions
\begin{align}
 I^{\rm fin}_{MB}\overset{d=4}{=} \frac{1}{16 \pi^2}\left[-1 +  2\log\left(\frac{m}{\mu}\right)
 + \frac{M^2- m^2 + s}{s}\log\left(\frac{M}{m}\right)\right. - \non
		   -\left.\frac{4p_{{\rm cms}} }{\sqrt{s} }\,\mathrm{arctanh}\left(\frac{2p_{{\rm cms}}\sqrt{s} }{(m + M)^2 - s}\right)\right]~,
\label{expli}
\end{align}
where $\mu$ is the regularization scale and $p_{{\rm cms}}$ is the modulus of the center-of-mass three momentum, expressed in terms of the
K\"all\'en function as $p_{{\rm cms}}=\lambda^{1/2}(s,m^2,M^2)/(2\sqrt{s})$. The $\mu$--dependence would be canceled by the corresponding scale
dependence of the higher-order counter terms. Dealing with a non-perturbative framework with only a finite number of terms being iterated, such
a cancellation is not possible, which is the reason why in most comparable approaches this scale is used as a free parameter. Here we fix it to
the values (in GeV) found in fitting strategy II of Ref.~\cite{Mai:2012wy}, namely $\log(\mu_\pi/(1 \text{GeV}))=-0.368$, $ \log(\mu_\eta/(1
\text{ GeV}))=0.056$ and $\log(\mu_K/(1 \text{ GeV}))=0.210$.

The following hadron masses and decay constants are used (all in MeV):
\begin{center}
\renewcommand{\arraystretch}{1.30}
\begin{tabular}{ l  l  l }
 $m_N=939$,            &$M_{\pi} =138$,   & $F_\pi  =92.4$,      \\
 $m_\Sigma=1195$,      &$M_{K}   =495$,   & $F_K    =113.0$,     \\
 $m_\Lambda=1115.7$,    &$M_\eta   =547$,  & $F_\eta =1.3F_\pi$.  \\
\end{tabular}\end{center}
The free parameters of the model are given by 14 low-energy constants of the next-to-leading chiral order, appearing in the potential $V$. All
14 parameters are adjusted here to reproduce the current SAID solution~\cite{Workman:2012jf} for the real and imaginary part of the $S_{11}$
partial wave in the energy region ${1080\le W\le 1800}$~MeV. The errors are assigned as described in Ref.~\cite{Bruns:2010sv}, namely $\Delta
S_{11}=0.005$ for $W\le 1280$~MeV and $\Delta S_{11}=0.030$ for higher energies. 

%%%%%%%%%%%%%%%%%%%%%%%%%%%%%%%%%%%%%%%%%%%%%%%%%%%%%%%%%%%%%%%%%%%%%%%%%%%%%%%%%%%%%%%%%%%
\begin{figure}
 \includegraphics[width=0.99\linewidth]{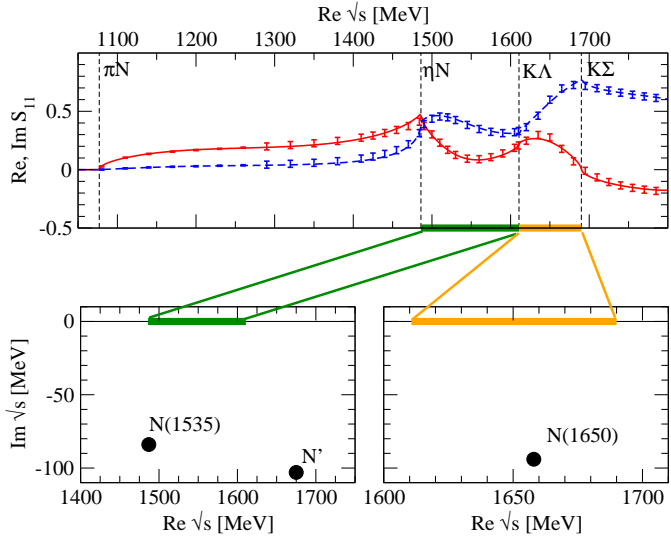}
 \caption{The $S_{11}$ amplitude. Upper panel: Best fit of the model to the energy independent solution of the PWA by the SAID group~\cite{Workman:2012jf}. Red (solid)
 and blue (dashed) lines represent the real and imaginary part of our solution, respectively, whereas the vertical dashed lines correspond to
 the two-particle thresholds.  Lower left panel: Riemann sheet connected to the physical axis between the $\eta N$ and the $K\Lambda$
 threshold. Right: sheet connected to the physical axis between $K\Lambda$ and $K\Sigma$ threshold. See text for the labeling of the poles.}
 \label{infVOL}
\vspace{-3mm}
\end{figure}
%%%%%%%%%%%%%%%%%%%%%%%%%%%%%%%%%%%%%%%%%%%%%%%%%%%%%%%%%%%%%%%%%%%%%%%%%%%%%%%%%%%%%%%%%%%
The best fit of our model is presented in Fig.~\ref{infVOL}. The fitted parameters are (all $b_i$ in GeV$^{-1}$):
\begin{align}
 b_1   &=-0.765, &b_6   &=-1.043,  &b_{11}&=-1.220,   \nonumber\\
 b_2   &=+0.924, &b_7   &=+5.919,  &b_0   &=-1.186,   \nonumber\\
 b_3   &=-2.610, &b_8   &=+0.732,  &b_D   &=+1.173,   \nonumber\\
 b_4   &=+0.892, &b_9   &=-1.304,  &b_F   &=-0.624,   \nonumber\\
 b_5   &=+0.023, &b_{10}&=+1.401.  & &                \nonumber
\end{align}
Note that these fit parameters differ from those in Ref.~\cite{Mai:2012wy} due to the on-shell approximation performed here.

At the pion-nucleon threshold we extract the scattering length to be $a_{\pi N}^{1/2}=1.13$~GeV$^{-1}$, which is somewhat smaller than
in the earlier analysis of Ref.~\cite{Mai:2012wy}, i.e. $a_{\pi N}^{1/2}=1.22$~GeV$^{-1}$.  The analytic structure of the amplitude is shown
in the two lower panels of Fig.~\ref{infVOL} for two sheets. The left panel shows the Riemann sheet that is directly connected to the physical
axis between the $\eta N$ and the $K\Lambda$ threshold as indicated with the thick horizontal bar. Performing the analytic continuation of the
scattering amplitude we locate the poles on this sheet at
\begin{align}
 W_{N(1535)}&= (1487 -i\,\,\,\, 84) \text{ MeV},\non 
 W_{N'~~~~~~}&= (1675 -i\, 103) \text{ MeV (hidden)},%\nonumber
\end{align}
where one pole can obviously be identified with the $N(1535)1/2^-$ resonance and another one ($N'$) is hidden behind the $K\Lambda$ threshold.
Apart from poles, we find on this sheet also a zero at $W_0=(1592-i\,62)$~MeV. The SAID group found a zero at
$W_0=(1578-i\,38)$~MeV~\cite{Arndt:2003if}. On the sheet connected to the physical axis between the $K\Lambda$ and the $K\Sigma$ threshold
(right panel), we find the pole of the $N(1650)1/2^-$ at
%\begin{align}
\begin{equation}
W_{N(1650)} = (1658 -i\, 94) \text{ MeV}.%\nonumber
\end{equation}
%\end{align}
Again, for the more precise determination of the pole positions and scattering lengths, using the full off-shell solution of the Bethe-Salpeter
equation, we refer the reader to original calculations \cite{Bruns:2010sv,Mai:2012wy}. In any case, the values found here are well within the
limits quoted by the Particle Data Group~\cite{Beringer:1900zz}.

We do not perform an error analysis on the extracted low energy constants or amplitudes as performed in Ref.~\cite{Mai:2012wy}. The error we are
interested in here is the one expected for the finite volume spectrum extrapolated to unphysical masses. This will be discussed below.

%%%%%%%%%%%%%%%%%%%%%%%%%%%%%%%%%%%%%%%%%%%%%%%%%%%%%%%%%%%%%%%%%%%%%%%%%%%%%%%%%%%%%%%%%%%

\subsection{\textbf{Discretization}}
\label{sec:diskre}
In the infinite volume all meson-baryon momenta are permitted while in a cubic volume of side length $L$ only momenta 
\be
\vec q=\frac{2\pi}{L}\,\vec n,\quad
\quad\vec n\in \mathds{Z}^3 
\ee
are allowed due to the imposed periodic boundary conditions. In particular, the momentum integration over intermediate meson-baryon states of
the propagator function $I_{MB}$ in Eq.~(\ref{scalar}) is replaced by a sum according to 
\be
\int\frac{d^3\vec q}{(2\pi)^3}\,f(|\vec q\,|)\to\frac{1}{L^3}\sum_{\vec n}\,f(|\vec q|) \ .
\label{substi}
\ee
To obtain the three-momentum part of the integration of $I_{MB}$ we integrate over the zero component with external momentum $p=(\sqrt{s},\vec
0)$,
\begin{align}
I_{MB}&=\int\frac{d^4q}{(2\pi)^4}\,\frac{i}{(p-q)^2-m+i\epsilon}\frac{1}{q^2-M^2+i\epsilon}\non
&=\int\frac{d^3\vec q}{(2\pi)^3}\,f(|\vec q|) \ , \non
&f(|\vec q|)=\frac{1}{2\omega_m(\vec q)\,\omega_M(\vec q)} \, \frac{\omega_m(\vec q)+\omega_M(\vec q)}
{s-\left[\omega_m(\vec q)+\omega_M(\vec q)\right]^2+i\epsilon} \ ,
\label{prop_cont}
\end{align}
where $\omega_m(\vec q)=\sqrt{m^2+\vec q^2}$ and $\omega_M(\vec q)=\sqrt{M^2+\vec q^2}$. Using Eq.~(\ref{substi}), this expression yields the
finite-volume propagator $\tilde I_{MB}$ that still requires a regularization. We can proceed similar to Ref.~\cite{Doring:2011vk} to express
$\tilde I_{MB}$ as
\be
\tilde I_{MB}=I_{MB}+\Delta I_{MB}
\ee
with the advantage that the regularization of the infinite volume is manifestly contained in $I_{MB}$, while $\Delta I_{MB}$ is the finite
difference between the infinite-volume and the finite-volume expression, given above threshold by
\begin{multline}
\Delta I_{MB}=\tilde I_{MB}(s)-I_{MB}(s)=\biggl\{\frac{1}{L^3}\sum_{\vec q}^{|\vec q|<q_{\rm max}}
-\int\limits^{|\vec q|<q_{\rm max}}\frac{d^3\vec q}{(2\pi)^3}\,\biggr\}
\\
\times\,\frac{1}{2\sqrt{s}}\frac{1}{p_{{\rm cms}}^2-\vec q^2+i\epsilon}
+\cdots
=\frac{1}{2\sqrt{s}}\,\frac{1}{L^3}\sum_{\vec q}^{|\vec q|<q_{\rm max}}
\frac{1}{p_{{\rm cms}}^2-\vec q^2}
\\
+\frac{1}{4\pi^2\sqrt{s}}\,
\left(q_{\rm max}+\frac{p_{{\rm cms}}}{2}\log\frac{q_{\rm max}-p_{{\rm cms}}}{q_{\rm max}+p_{{\rm cms}}}\right)
+\frac{ip_{{\rm cms}}}{8\pi\,\sqrt{s}}+\cdots\, ,
\label{eq:equiv}
\end{multline}
where the ellipses stand for the exponentially suppressed terms.  Below threshold, the last term on the r.h.s. becomes real as the analytic
continuation of $p_{{\rm cms}}$ becomes imaginary.

Moreover, as seen from Eq.~(\ref{eq:equiv}), one may in fact remove here the cutoff, sending $q_{\rm max}\to\infty$.  Indeed, one should
obviously take a $q_{\rm max}$ such that $p_{{\rm cms}}^2< q_{\rm max}^2$ in the whole region of interest to us. If we sum and integrate from
$q_{\rm max}$ to $q_{\rm max}'$, with $q_{\rm max}'>q_{\rm max}$, the denominator $p_{{\rm cms}}^2-\vec q^2$ is  not singular and, according to
the regular summation theorem, only exponentially suppressed corrections may arise. Finally, noting that (see, e.g. Ref.~\cite{beane})
\begin{multline}
\label{eq:Z00}
\lim_{q_{\rm max}\to\infty}\biggl\{\frac{1}{L^3}
\sum_{\vec q}^{|\vec q|<q_{\rm max}}
\frac{1}{p_{{\rm cms}}^2-\vec q^2}-\frac{q_{\rm max}}{2\pi^2}\biggr\}
%=-\frac{1}{2\pi^{3/2}L}\,{\cal Z}_{00}(1,\hat p^2)\, %\nonumber
=-\displaystyle\frac{{\cal Z}_{00}(1,\hat p^2)}{2\pi^{3/2}L}\, , %\nonumber
\end{multline}
where $\hat p=({\vec pL})/({2\pi})$ and ${\cal Z}_{00}$ stands for the L\"uscher zeta-function~\cite{luscher,Luscher:1990ux}, we can identify
$\tilde I_{MB}$ with the L\"uscher function up to exponentially suppressed terms
\ba
\tilde I_{MB}\simeq I_{MB}-\frac{1}{4\pi^{3/2}\sqrt{s}L}\,{\cal Z}_{00}(1;\hat p^2)
+i\,\frac{p_{{\rm cms}}}{8\pi \sqrt{s}} \ . %\nonumber
\label{rela1}
\ea
In practical terms, we obtain the finite volume propagator by substituting the imaginary part of the infinite volume propagator according to
\be
\tilde I_{MB}={\rm Re}\,I^{\rm fin}_{MB}+\delta I_{MB}^{},\quad
\delta I_{MB}=\tilde G_S-{\rm Re}\,G_S
\label{practical}
\ee
with $\tilde G_S$ and $G_S$ defined in Eq.~(15) of Ref.~\cite{Doring:2011vk} and $I_{MB}^{\rm fin}$ from Eq.~(\ref{expli}). The summation over
lattice momenta can be simplified by the use of the $\theta$--series~\cite{Doring:2011ip}. It should also be stressed that up to exponentially
suppressed terms this is equivalent to the $K$-matrix formalism developed in Ref.~\cite{Bernard:2010fp}. A very similar approach to evaluate the
discretized version of dimensionally regularized loops has been developed in Ref.~\cite{MartinezTorres:2012yi}.

Hybrid boundary conditions were introduced in Ref.~\cite{Okiharu:2005eg} to distinguish scattering states from tightly bound quark-antiquark
systems. Similarly, as proposed in Refs.~\cite{Bernard:2010fp,Doring:2011vk}, twisted boundary conditions provide the possibility to change
thres\-holds in lattice gauge calculations. This provides a unique opportunity to study the nature of resonances that lie close to a threshold 
like, for example, the $f_0(980)$ with regard to the $\bar KK$ threshold~\cite{Morgan:1993td,Baru:2004xg},  because the twisting moves the
threshold while the resonance stays put.   We realize that it could be quite challenging to implement this idea (including twisting for the sea
quarks) in present-day lattice simulations. 

In chiral unitary approaches, the $N(1535)$ and $N(1650)$ resonances exhibit a very strong (sub)threshold coupling to the $K\Lambda$ and
$K\Sigma$ channels. In fact, the $N(1535)$ is often seen as a quasibound $KY$ state in that picture~\cite{Kaiser:1995cy}. If one imposes
different boundary conditions on the strange quark than on up and down quarks, one expects a strong response of these resonances to the modified
boundary conditions. This would be in contrast to the picture in which these resonances couple only moderately to the $KY$-channels. In that
case, a modification of the boundary conditions would only have minor impact.

With this idea in mind, we formulate the discretization for maximally twisted, i.e. antiperiodic boundary conditions.  Twisted  boundary
conditions for the strange quark have been introduced in Ref.~\cite{Bernard:2010fp}, $s(\vec x+L\,\hat e_i)=e^{i\theta_i}s(\vec x)$ where the
$\hat e_i$ are the unit vectors along the lattice axes and $0\leq \theta_i<2\pi$.  If the up and down quarks remain with periodic boundary
conditions, i.e. $u(\vec x+L\,\hat e_i)=u(\vec x)$, $d(\vec x+L\,\hat e_i)=d(\vec x)$, the twisting angle appears only in the $K$, $\Lambda$,
and $\Sigma$ fields, but not in the $\pi$, $\eta$, and $N$ fields,
\begin{align}
K^\pm(\vec x+L\,\hat e_i)	&=e^{\mp i\theta_i} K^\pm(\vec x), 	&K^0(\vec x+L\,\hat e_i)		&=e^{-i\theta_i}K^0(\vec x), 		\non
\bar K^0(\vec x+L\,\hat e_i)	&=e^{i\theta_i}\bar K^0(\vec x), 	&\Sigma^{\pm}(\vec x+L\,\hat e_i)	&=e^{i\theta_i} \Sigma^{\pm}(\vec x), 	\non
\Sigma^{0}(\vec x+L\,\hat e_i)	&=e^{i\theta_i} \Sigma^{0}(\vec x),	&\Lambda(\vec x+L\,\hat e_i)		&=e^{i\theta_i} \Lambda,
\end{align}
effectively leading to a change in the summation~\cite{Bernard:2010fp} over the lattice momenta of the $KY$ channels,  
\be
\sum_{\vec n}\,f\left(\left|\vec q=\frac{2\pi}{L}\,\vec n\right|\right)\to\sum_{\vec n}\,f\left(\left|\vec q=\frac{2\pi}{L}\,\vec n+\frac{\vec \theta}{L}\right|\right) \ ,
\label{anti}
\ee
where $\vec \theta$ is the twisting angle and antiperiodic boundary conditions in all three space dimensions correspond to
$\vec\theta=(\pi,\pi,\pi)$. The summations for the $\pi N$ and $\eta N$ channels are not affected. Using Eq.~(\ref{anti}) for the $f$ of
Eq.~(\ref{prop_cont}) it is straightforward to obtain the antiperiodic finite volume propagator from Eq.~(\ref{practical}). The summation for
antiperiodic boundary conditions can be simplified by using properties  of the elliptic $\vartheta_2$-function as derived in
Ref.~\cite{Doring:2011ip}.

The eigenlevels in the finite volume are given by the poles of the solution $\tilde T$  of the coupled-channel scattering equation 
\be
\tilde T=V^{\rm on}+V^{\rm on}\tilde G\tilde T,\quad \tilde G=\left(\slashed{p}\frac{p^2-M^2+m^2}{2 p^2}+m\right)\,\tilde I_{MB}
\ee 
with $\tilde I_{MB}$ from Eq.~(\ref{practical}). The finite volume effects arise, thus, entirely from the modified propagator $\tilde G$ in the
various channels while the contact interactions $V^{\rm on}$ remain unchanged.

Rotational symmetry is broken in the finite volume. As it is well known~\cite{luscher,Luscher:1990ux}, the $S$-wave amplitude considered here
mixes with $G$-wave amplitudes. We neglect this effect because the centrifugal barrier effectively suppresses the $G$-wave amplitude up to the
considered energies. In principle, there are many more open channels that are neglected in this work, starting at the $\pi\pi N$ threshold. A
(still incomplete) coupling scheme for $J^P=1/2^-$ can be seen in Table IX of Ref.~\cite{Ronchen:2012eg}. Those effects are relevant especially
in the meson-baryon sector~\cite{Doring:2009yv,Doring:2010ap,Ronchen:2012eg}, but in the $S_{11}$ partial wave the inelasticities are dominated
by the $\eta N$ channel and effects from $\pi\pi N$ and other multi-meson-states are neglected in this exploratory study. Pioneering work to
study, at least in principle, three-body systems in the finite volume have emerged
recently~\cite{Kreuzer:2010ti,Polejaeva:2012ut,Roca:2012rx,Guo:2012hv,Briceno:2012rv}.

In the present work, we restrict ourselves to the prediction and study of lattice levels in the overall center-of-mass frame. Moving frames
provide additional levels at different scattering energies and are nowadays a standard tool in lattice
calculations~\cite{Rummukainen:1995vs,Bour:2011ef,Davoudi:2011md,Fu:2011xz,Pelissier:2011ib,Leskovec:2012gb,Dudek:2012gj,Briceno:2012yi}. 
The extension of the
present formalism to moving frames is in principle straightforward and has been worked out in Ref.~\cite{Doring:2012eu} although it has to be
stressed that the group structure of the spin-$\nicefrac{1}{2}$ spin-0 system is slightly different~\cite{Gockeler:2012yj}, let alone the fact that other channels (e.g., $\rho N$) couple with different angular momenta to the $J^P=1/2^-$ sector~\cite{Ronchen:2012eg}.

%%%%%%%%%%%%%%%%%%%%%%%%%%%%%%%%%%%%%%%%%%%%%%%%%%%%%%%%%%%%%%%%%%%%%%%%%%%%%%%%%%%%%%%%%%%

\section{Results}

The prediction of the energy levels on any specific lattice requires the knowledge of the meson and baryon masses as well as the meson decay
constants calculated on this lattice. We will rely here on two different parameter sets, determined by the European Twisted Mass (ETMC) and the
QCDSF collaborations.

\subsection{\textbf{Set A - ETMC}}

%%%%%%%%%%%%%%%%%%%%%%%%%%%%%%%%%%%%%%%%%%%%%%%%%%%%%%%%%%%%%%%%%%%%%%%%%%%%%%%%%%%%%%%%%%%

In this setup the meson masses and pion decay constant are taken from the recent calculation in $N_f=2+1+1$ twisted mass lattice QCD, i.e.
ensemble $B25.32$ of Ref.~\cite{Ottnad:2012fv}. For the lattice size of $L/a=32$ and spacing $a=0.078$~fm, the pion mass is fixed there to
$M_\pi=269$~MeV, whereas the strange quark mass is held approximately at the physical value. As the kaon and eta decay constants are not
available in this calculation at the moment, we decide to relate them to $F_\pi$ with typical ratios of $1.15$ and $1.3$, respectively. The
baryon masses are also taken from a calculation by the ETM collaboration,
however, with only two dynamical quarks and
an older lattice action, see
Ref.~\cite{Alexandrou:2009qu}. Nevertheless, the strange quark mass is held again approximately at the physical value and $M_\pi=269$~MeV for
the identical lattice size and comparable lattice spacing, i.e. $a=0.0855$~fm. Altogether, the assumed parameters in the finite volume read in
MeV:
\begin{center}
\renewcommand{\arraystretch}{1.30}
\begin{tabular}{ l  l  l }
 $m_N^{{\rm Set~A}}=1142$,	  &$M_{\pi~}^{{\rm Set~A}} =269$,  & $F_\pi^{{\rm Set~A}}  =102.1$,  \\ 
 $m_\Sigma^{{\rm Set~A}}=1359$,    &$M_{K~}^{{\rm Set~A}}   =535$,  & $F_K^{{\rm Set~A}}    =117.4$,  \\  
 $m_\Lambda^{{\rm Set~A}}=1295$,   &$M_\eta^{{\rm Set~A}}   =589$,  & $F_\eta^{{\rm Set~A}} =132.7$.  \\
\end{tabular}\end{center}
With these parameters we first discuss the infinite-volume quantities. The scattering length reads $a_{\pi N}^{1/2}=0.73$ GeV$^{-1}$ and is
around 35\% smaller than the one at the physical point. This is due to the NLO terms, which become quite large already at the $\pi N$
threshold. 

%%%%%%%%%%%%%%%%%%%%%%%%%%%%%%%%%%%%%%%%%%%%%%%%%%%%%%%%%%%%%%%%%%%%%%%%%%%%%%%%%%%%%%%%%%%
\begin{figure}
 \includegraphics[width=0.99\linewidth]{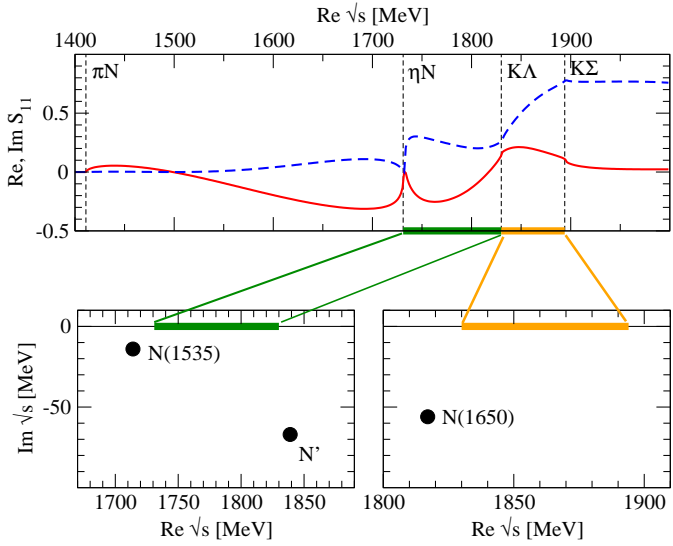}
 \caption{Upper panel: Real (solid line) and imaginary part (dashed line) of the $S_{11}$ amplitude, chirally extrapolated using masses and decay
 constants of the ETM collaboration. Lower panels: two of the Riemann sheets with poles. Labeling as in Fig.~\ref{infVOL}.}
 \label{fig:etmc}
\vspace{-3mm}
\end{figure}
%%%%%%%%%%%%%%%%%%%%%%%%%%%%%%%%%%%%%%%%%%%%%%%%%%%%%%%%%%%%%%%%%%%%%%%%%%%%%%%%%%%%%%%%%%%
The $S_{11}$ amplitude, with the masses and decay constants of the ETM collaboration, is shown in the upper panel of Fig.~\ref{fig:etmc}.
Comparing to Fig.~\ref{infVOL}, all thresholds have moved to higher energies. The cusp at the $\eta N$ threshold has become more pronounced, but
no clear resonance shapes are visible. The structure of the amplitude becomes clearer by inspecting the complex energy plane on different
Riemann sheets. This is visualized in the lower panels of Fig.~\ref{fig:etmc}. The Riemann sheets and labeling of the poles are the same as in
Fig.~\ref{infVOL}.  The pole positions are
\begin{align}
 W_{N(1535)}^{\rm Set~A}&= (1714 -i\, 14) \text{ MeV (hidden)}, \non
 W_{N'~~~~~~~}^{\rm Set~A}&= (1839 -i\, 67) \text{ MeV (hidden)}, \non
 W_{N(1650)}^{\rm Set~A}&= (1817 -i\, 56) \text{ MeV (hidden)}.%\nonumber
\end{align}
Compared to the physical point, the imaginary parts of the pole positions became much smaller due to the reduced phase space. Both the
thresholds and the real parts of the pole positions have moved to higher energies. However, the thresholds have moved farther than the pole
positions, such that the $N(1535)$ and $N(1650)$ poles are no longer situated below the part of the respective sheet, that is connected to the
physical axis (thick horizontal lines). The poles are thus hidden and no clear resonance signals are visible in the physical amplitude.
Instead, the amplitude is dominated by cusp effects. 

%%%%%%%%%%%%%%%%%%%%%%%%%%%%%%%%%%%%%%%%%%%%%%%%%%%%%%%%%%%%%%%%%%%%%%%%%%%%%%%%%%%%%%%%%%%
\begin{figure}
 \includegraphics[width=0.9\linewidth]{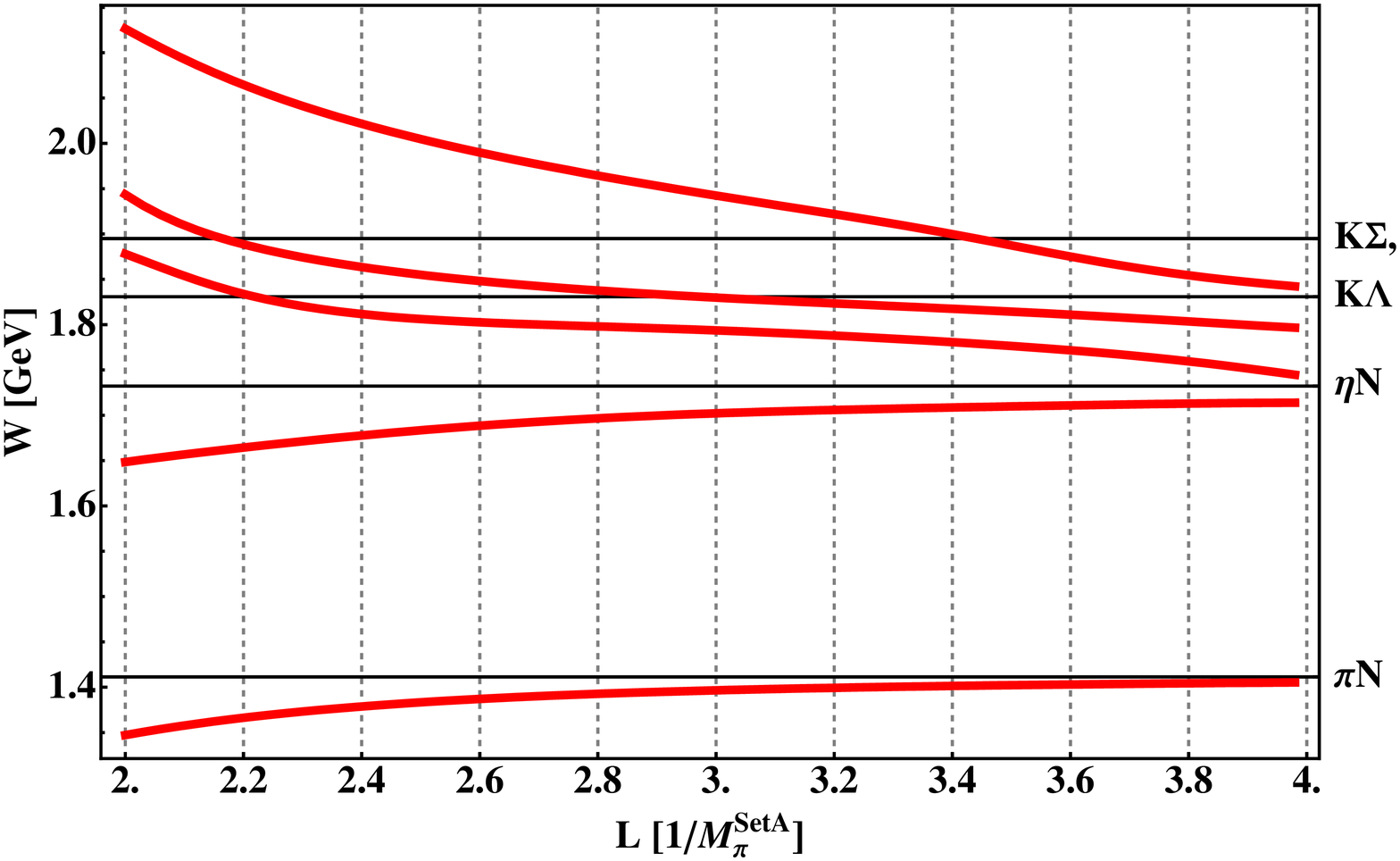}
 \includegraphics[width=0.9\linewidth]{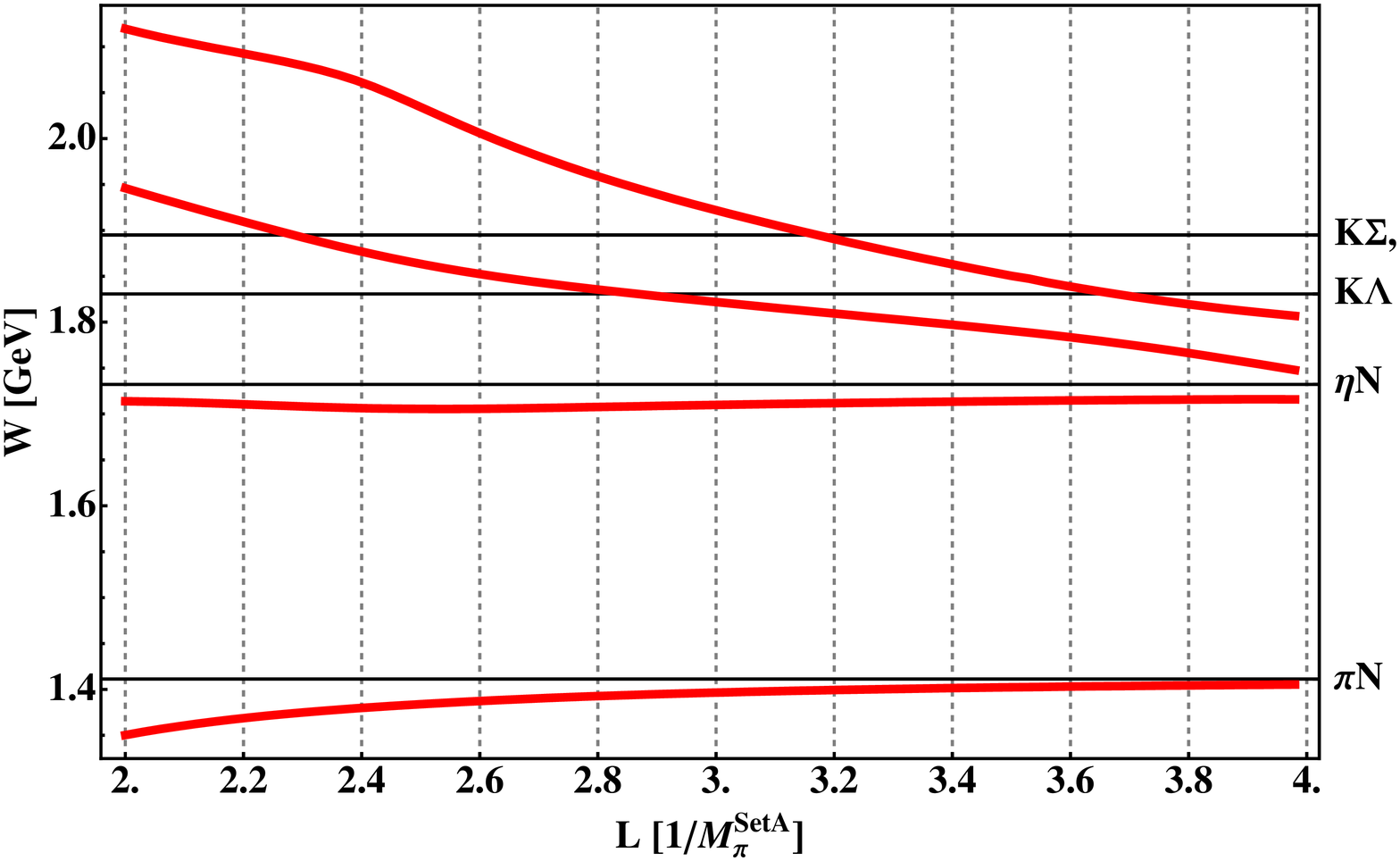}
 \caption{Volume dependence of the energy levels predicted by our model for $I=1/2$ $\pi N$ scattering for periodic (upper) and antiperiodic
(lower) boundary conditions. Masses and pion decay constant are taken from ETMC, see
Refs.~\cite{Ottnad:2012fv,Alexandrou:2009qu}.}
 \label{finVOL_ETM}
\vspace{-3mm}
\end{figure}
%%%%%%%%%%%%%%%%%%%%%%%%%%%%%%%%%%%%%%%%%%%%%%%%%%%%%%%%%%%%%%%%%%%%%%%%%%%%%%%%%%%%%%%%%%%
Having analyzed the infinite-volume solution, we now turn to the finite-volume spectrum. By discretizing the model as outlined in
Sec.~\ref{sec:diskre} we obtain the prediction for the volume dependence of the energy levels shown in Fig.~\ref{finVOL_ETM}. In the upper
panel, the spectrum for periodic boundary conditions is shown. The lowest level at the $\pi N$ threshold exhibits the characteristic $1/L^3$
dependence that can serve to calculate the (attractive) scattering length. 

Similarly, the next level is situated close to the $\eta N$ threshold. This level is not induced by the presence of a resonance but a genuine
effect of an $S$-wave threshold in a multi-channel problem. Such inelastic thresholds induce the same avoided level crossing as resonances,
discussed in detail in Refs.~\cite{Bernard:2010fp,Doring:2011vk}. One striking example discussed there is the one of the $f_0(980)$ close to the $\bar KK$
thresholds: irrespectively of whether the resonance is present or not, there is avoided level crossing, and the levels are only slightly shifted
if the resonance, albeit being so narrow, is present. Thus, the level below the $\eta N$ threshold shown in Fig.~\ref{finVOL_ETM} cannot be
attributed to the $N(1535)$ resonance. In any case, the $N(1535)$ is on a different sheet and, moreover, hidden as discussed following
Fig.~\ref{fig:etmc}. 

The following two levels, beyond the $\eta N$ threshold, are more difficult to interpret. They both show a plateau that, however, cannot be
uniquely attributed neither to the $KY$ threshold nor to the hidden resonances. The rather involved interplay between hidden poles and
threshold openings hinders the straightforward extraction of resonances. 

The lower panel of Fig.~\ref{finVOL_ETM} shows the level spectrum if antiperiodic boundary conditions are applied to the strange quark. As
discussed in Sec.~\ref{sec:diskre}, this results in unchanged propagators for the $\pi N$ and $\eta N$ channels, while the $KY$ channels undergo
modifications. In particular, the summation over lattice momenta is shifted from the origin, resulting in a finite relative momentum for the
$KY$ pair at rest, of $\pm (\pi/L,\pi/L,\pi/L)$~\cite{Bernard:2010fp}. Accordingly, the singularities at the $KY$ thresholds, induced by the
denominator of $f$ in Eq.~(\ref{prop_cont}), are shifted and the avoided level crossing associated with $S$-wave thresholds disappears. Indeed,
the third and fourth level, that showed avoided crossing with periodic boundary conditions, have a very different $L$-dependence with
antiperiodic boundary conditions as Fig.~\ref{finVOL_ETM} shows. In particular, the plateaus have disappeared.

Even for the second level, below the $\eta N$ threshold, we observe small changes of the $L$-dependence, even though the boundary conditions are
only changed for the higher-lying $KY$ channels. This demonstrates that changes of the boundary conditions for the strange quark can indeed have
an effect for the sub-threshold dynamics. 

In summary, the resonance poles for the ETMC setup lie on hidden sheets. Plateaus of the $L$-dependence of the levels are rather tied to
two-particle thresholds than to resonances, and neither with periodic nor antiperiodic boundary conditions a direct access to the $N(1535)$ or
$N(1650)$ resonances is possible. In Sec.~\ref{sec:discu} we discuss strategies how to proceed in such a case.

%%%%%%%%%%%%%%%%%%%%%%%%%%%%%%%%%%%%%%%%%%%%%%%%%%%%%%%%%%%%%%%%%%%%%%%%%%%%%%%%%%%%%%%%%%%

\subsection{\textbf{Set B - QCDSF}}

%%%%%%%%%%%%%%%%%%%%%%%%%%%%%%%%%%%%%%%%%%%%%%%%%%%%%%%%%%%%%%%%%%%%%%%%%%%%%%%%%%%%%%%%%%%
\begin{figure}
 \includegraphics[width=0.99\linewidth]{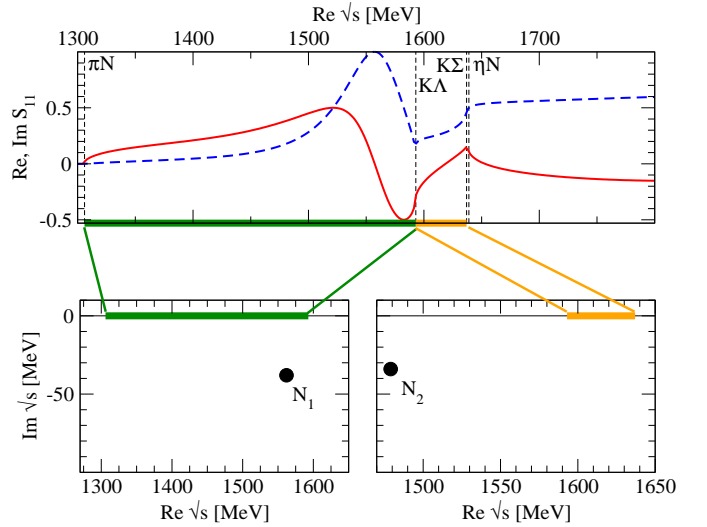}
 \caption{Upper panel: Real (solid line) and imaginary part (dashed line) of the $S_{11}$ amplitude, chirally extrapolated using masses and decay
constants of the QCDSF collaboration~\cite{Bietenholz:2011qq}. Lower panels: two of the Riemann sheets with poles. Labeling as in
Fig.~\ref{infVOL}.}
 \label{fig:qcdsf}
\vspace{-3mm}
\end{figure}
%%%%%%%%%%%%%%%%%%%%%%%%%%%%%%%%%%%%%%%%%%%%%%%%%%%%%%%%%%%%%%%%%%%%%%%%%%%%%%%%%%%%%%%%%%%

For the second set of parameters we choose a setup employed by the QCDSF collaboration~\cite{Bietenholz:2011qq}. Here, baryon and meson masses
are determined from an alternative approach to tune the quark masses. Most importantly, while the lattice size and spacing are comparable to
those of the ETMC, i.e. $L/a=32$ and $a=0.075$~fm, the strange quark mass differs significantly from the physical value. The latter results in a
different ordering of the masses of the ground-state octet mesons and, consequently, in a different ordering of meson-baryon thresholds. For
further details we refer the interested reader to Ref.~\cite{Bietenholz:2011qq}. Altogether, the lattice input for our calculation reads
\begin{center}
\renewcommand{\arraystretch}{1.20}
\begin{tabular}{ l  l  l }
 $m_N^{\rm Set~B}=1020$,  &$M_{\pi~}^{\rm Set~B} =286$,  & $F_\pi^{\rm Set~B}  =106.6$,  \\
 $m_\Sigma^{\rm Set~B}=1155$,  &$M_{K~}^{\rm Set~B}   =482$,  & $F_K^{\rm Set~B}    =115.3$,  \\
 $m_\Lambda^{\rm Set~B}=1111$,  &$M_\eta^{\rm Set~B}   =619$,  & $F_\eta^{\rm Set~B} =127.6$,  \\
\end{tabular}
\end{center}
where the meson decay constants have been calculated from next-to-leading
order chiral perturbation theory. Two low-energy
constants enter the calculation, for which the world lattice results were taken from \cite{Colangelo:2010et}, i.e. $L_4=0.04$ and $L_5=0.84$.

As discussed before, the NLO contributions become quite large already at energies slightly above the $\pi N$ threshold, adding up destructively
with the contribution from the Weinberg-Tomozawa term. This reduces the scattering length compared to the one calculated with the physical
parameters, like in the ETMC case. However, for the QCDSF values the $K\Sigma$ threshold lies closer to the $\pi N$ threshold than in the
physical or in the ETMC set. Consequently, the $K\Sigma$ loops contribute stronger to the pion-nucleon scattering amplitudes, which yields an
overall pion-nucleon scattering length of $a_{\pi N}^{1/2}=1.18$ GeV$^{-1}$. This is almost the size of the physical value and larger than in
the ETMC set discussed before. 

The scattering length depends on the input masses and decay constants, let alone the model dependence. However, for none of our parameter
configurations we observe such a large $\pi N$ scattering length as reported
by Lang and Verduci \cite{Lang:2012db}, their value being
$5.3\pm 1.4$~GeV$^{-1}$.

\begin{figure}
 \includegraphics[width=0.9\linewidth]{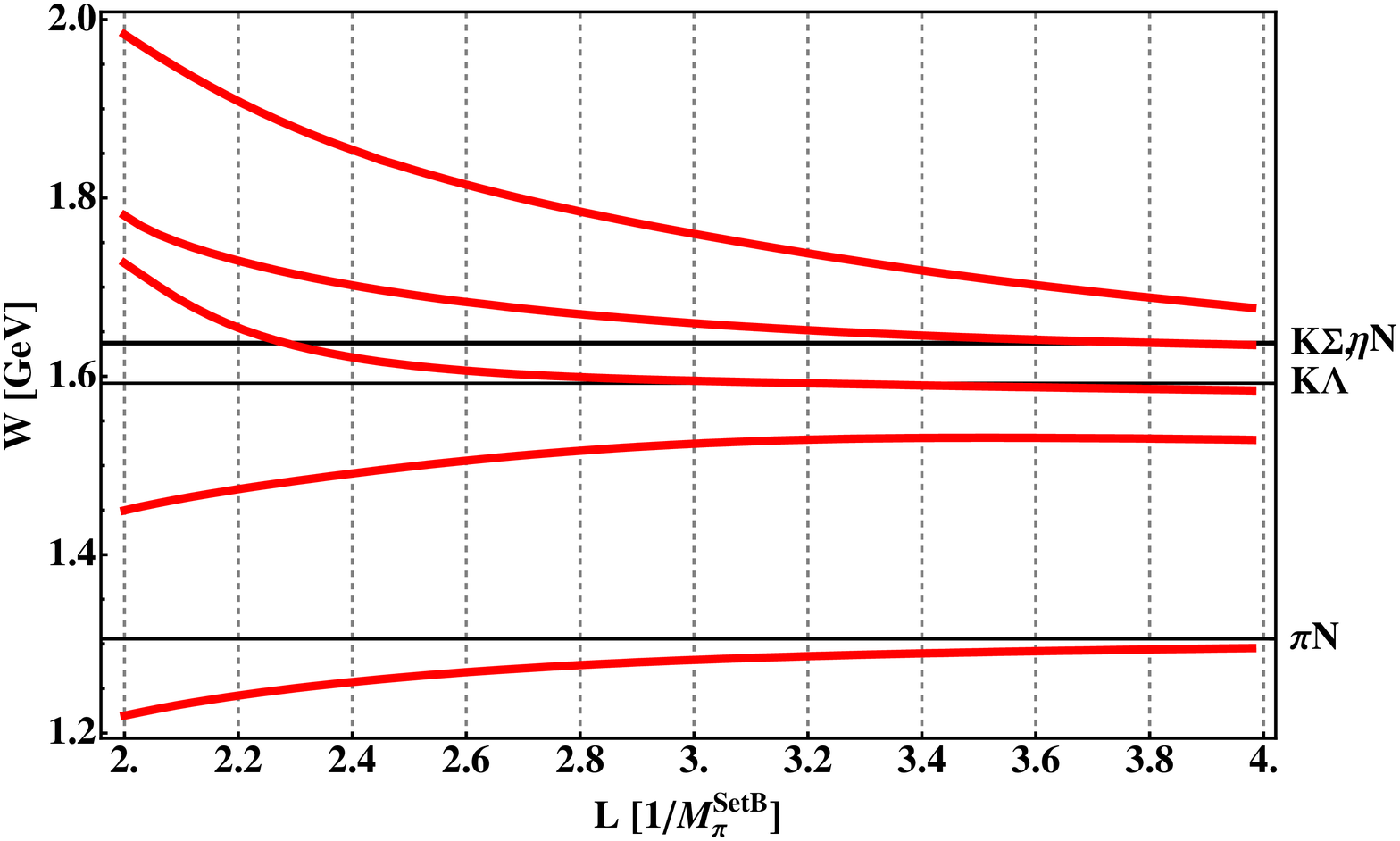}
 \includegraphics[width=0.9\linewidth]{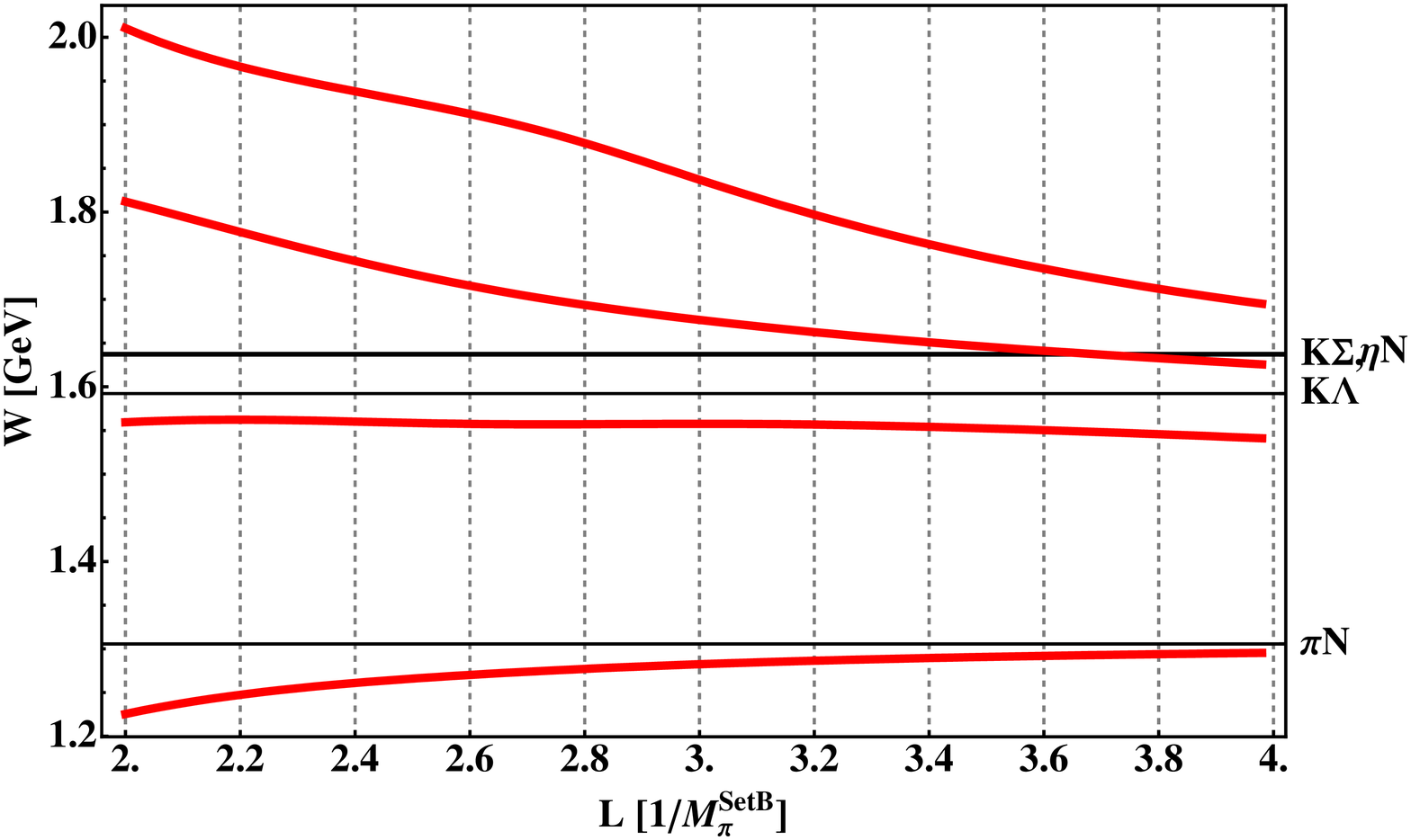}
 \caption{Volume dependence of the energy levels predicted by our model for $I=1/2$ $\pi N$ scattering for periodic (upper) and antiperiodic
(lower) boundary conditions. Masses and decay constants are taken from the calculation by the QCDSF collaboration, see
Ref.~\cite{Bietenholz:2011qq}.}
 \label{finVOL_QCDSF}
\vspace{-3mm}
\end{figure}

The amplitude using the QCDSF parameter set is shown in the upper panel of Fig.~\ref{fig:qcdsf}. In contrast to the ETMC case, a clear resonance
signal is visible below the $K\Lambda$ threshold, that is the first inelastic channel in this parameter setup. Indeed, we find a pole $N_1$ on
the corresponding Riemann sheet, as indicated in the  lower left panel. Unlike in the ETMC case, it is not hidden behind a threshold. Between
the $K\Lambda$ and the $K\Sigma$ threshold, there is only the hidden pole $N_2$ (right panel). The $K\Sigma$ and $\eta N$ thresholds are almost
degenerate and on sheets corresponding to these higher-lying thresholds we only find hidden poles.
The precise pole positions are
\begin{align}
 &W_{N_1}^{\rm Set~B}=      (1562 -    i\, 38) \text{ MeV}, \non
 &W_{N_2}^{\rm Set~B}= (1479 -i\, 34) \text{ MeV (hidden)}. %\nonumber
% &\sqrt{s}_{K\Sigma-\eta N}^{\rm QCDSF}= (1583 -i\, 89) \text{ MeV (hidden)}. \nonumber
\end{align}
Discretizing the present model as described in Sec.~\ref{sec:diskre} we obtain the energy levels shown in Fig.~\ref{finVOL_QCDSF}. 
%%%%%%%%%%%%%%%%%%%%%%%%%%%%%%%%%%%%%%%%%%%%%%%%%%%%%%%%%%%%%%%%%%%%%%%%%%%%%%%%%%%%%%%%%%%
%%%%%%%%%%%%%%%%%%%%%%%%%%%%%%%%%%%%%%%%%%%%%%%%%%%%%%%%%%%%%%%%%%%%%%%%%%%%%%%%%%%%%%%%%%%
For periodic boundary conditions (upper panel) we observe that most of energy levels are close to the two-particle thresholds for large $L$. At
the position of the $N_1$ pole, there is a level. However, we have here a pole close to a threshold, with a second channel open. This is
precisely the situation of the $f_0(980)$ discussed in depth in Ref.~\cite{Doring:2011vk}. In the following section we discuss strategies to
extract the amplitude in such cases. 

Using antiperiodic boundary conditions for the strange quark, as shown in the lower panel of Fig.~\ref{finVOL_QCDSF}, the singularity at the
$K\Lambda$ threshold is removed. In that case, we observe an almost $L$-independent level very close to the position of the $N_1$ pole. As the
resonance is quite narrow, one might identify this level with the $N_1$ pole, although there are, of course, still finite-volume corrections. 

While the extraction of the $N_1$ pole is quite promising for the parameters used by the QCDSF collaboration, one should still realize that this
pole is on a different sheet than the one of the $N(1535)$ or the
$N(1650)$. It is not evident what happens to this pole
 as the masses and decay constants approach the physical point and thus the
various thresholds get ordered correctly.

%%%%%%%%%%%%%%%%%%%%%%%%%%%%%%%%%%%%%%%%%%%%%%%%%%%%%%%%%%%%%%%%%%%%%%%%%%%%%%%%%%%%%%%%%%%

\subsection{\textbf{Discussion and outlook}}
\label{sec:discu}
The negative parity $S_{11}$ partial wave in meson-baryon scattering is very complex due to many threshold openings and two resonances, one of
which strongly coupling to the $\eta N$ channel. Those resonances are also believed to have a strong sub-threshold couplings to the $KY$
channels. Fitting the physical amplitude and thus fixing low-energy constants and scales, we use the quark mass dependence of the
Weinberg-Tomozawa and the NLO contact terms to predict the amplitude for typical lattice setups. Depending on the masses and decay constants,
resonance poles may become hidden behind thresholds as in case of the ETMC setup. In the QCDSF setup, there is one pole visible as a prominent
resonance on the physical. However, in that setup the threshold
ordering is reversed and it is not clear what happens to this pole as the
quark masses are lowered.

For these amplitudes at unphysical quark masses, we have predicted the finite-volume level-spectrum. Resonances usually manifest themselves in
avoided level crossing. However, in $S$-wave there is the additional complication that inelastic thresholds induce the same pattern. If
resonances are close to thresholds, it is very difficult to disentangle the dynamics, as is observed for both setups studied. The effect of the
$KY$ thresholds may be reduced by introducing twisted boundary conditions for the strange. Indeed, for the QCDSF setup we observe an almost
$L$-independent level close to the resonance position. This shows that changing the boundary conditions promises indeed for a cleaner resonance
extraction, although the technical realization on the lattice is intricate.  In any case, modified boundary conditions for the strange quark
shed light on the nature of the $J^P=1/2^-$ resonances and their supposed strong coupling to the hidden strangeness $KY$ channels.

In Ref.~\cite{Doring:2011vk} it was discussed how to combine lattice data from different boundary conditions to extrapolate resonances in a
two-channel problem to the infinite-volume limit. If one makes minimal assumptions on the smoothness of the potential, information from
different energies may be combined to allow for a quantitative resonance extraction. A complimentary way to obtain more information from the
lattice, without having to change the volume, is the use of moving frames. Unlike the $\pi\pi$ case, in meson-baryon scattering there are,
however, many large higher partial waves of different parity, and the disentanglement of the $S$-wave contribution might become difficult. 

In summary, we have shown that due to the many thresholds the $N(1535)$ and $N(1650)$ may become hidden in a unitary chiral extrapolation of the
amplitude to unphysical quark masses. The extrapolation to the infinite volume poses additional problems due to a complicated
threshold-resonance interplay requiring special techniques as modified boundary conditions to disentangle the resonance dynamics.

\noindent
\textbf{Acknowledgments}
We thank A.~Rusetsky for useful discussions. 
This work is supported in part by the EU Integrated Infrastructure Initiative
HadronPhysics3 and the 
DFG  through funds provided to the Sino-German CRC~110 ``Symmetries and the
Emergence of Structure in QCD'' and to the CRC~16 ``Subnuclear Structure of
Matter''. 

\vspace*{-0.2cm}
%%%%%%%%%%%%%%%%%%%%%%%%%%%%%%%%%%%%%%%%%%%%%%%%%%%%%%%%%%%%%%%%%%%%%%%%%%%%%%%%%%%%%%%%%%%

\end{document}